# Constructing a Social Accounting Matrix Framework to Analyse the Impact of Public Expenditure on Income Distribution in Malaysia

## *(Pembinaan Rangka Kerja Matrik Perakaunan Sosial bagi Menganalisis Kesan Perbelanjaan Awam ke atas Pengagihan Pendapatan di Malaysia)*

**Mukaramah-Harun**
Universiti Utara Malaysia

**A.R. Zakariah**
Malaysian Institute of Economic Research

**M. Azali**
Universiti Putra Malaysia

*ABSTRACT*

*The use of the social accounting matrix (SAM) in income distribution analysis is a method recommended by economists. However, until now, there have only been a few SAM developed in Malaysia. The last SAM produced for Malaysia was developed in 1984 based upon data from 1970 and has not been updated since this time despite the significance changes in the structure of the Malaysian economy. The paper proposes a new Malaysian SAM framework to analyse public expenditure impact on income distribution in Malaysia.The SAM developed in the present paper is based on more recent data, providing an up-to date and coherent picture of the complexity of the Malaysian economy.The paper describes the structure of the SAM framework with a detailed aggregation and disaggregation of accounts related to public expenditure and income distribution issues. In the SAM utilized in the present study, the detailed framework of the different components of public expenditure in the production sectors and household groups is essential in the analysis of the different effects of the various public expenditure programmes on the incomes of households among different groups. The SAM utilized in the present study can be applied to answer questions concerning whether components of public expenditure expansion would benefit the poor and, if so, which components are most likely to be beneficial. Other accounts are in aggregate form, such as the accounts for companies, accounts for private capital investment, and accounts for the Rest of the World (ROW). When evaluating public sector expenditure impacts on income distribution, most economists use a variety of policy tools including econometric models, cost-effectiveness analysis and social cost benefit analysis. These existing tools rely on estimation procedures that do not account for the complex interactions between poverty; income distribution; the endogeneity of the income distributions; and other variables. The strength of the SAM utilized in the present study over the tools mentioned above lies in the consistency of the modelling of the distribution of income and its ability to trace out chains of linkages from changes in demand to changes in production, factor incomes, household incomes and final demands.*

*ABSTRAK*

*Penggunaan SAM dalam analisis agihan pendapatan merupakan kaedah yang selalu disarankan oleh ahli ekonomi, walau bagaimanapun sehingga kini hanya terdapat beberapa SAM yang telah dibentuk di Malaysia. SAM yang terkini diterbitkan pada tahun 1984 dengan menggunakan data tahun 1970 dan sejak dari itu ia tidak dikemaskini walaupun berlaku perubahan ketara dalam struktur ekonomi. Kertas kerja ini membentangkan pembinaan SAM Malaysia untuk menganalisis kesan perbelanjaan awam ke atas agihan pendapatan di Malaysia. Oleh itu SAM yang dibina di sini adalah berdasarkan kepada data terbaru, menyediakan gambaran terkini yang lebih jelas terhadap kerumitan ekonomi. Kertas kerja ini akan menerangkan struktur SAM dengan akaun agregat dan akaun yang tidak diagregat berkaitan dengan isu-isu perbelanjaan awam dan pengagihan pendapatan. Dalam SAM, rangka kerja yang terperinci bagi komponen yang berbeza dalam perbelanjaan awam di sektor pengeluaran dan kumpulan isi rumah yang berbeza adalah penting untuk menganalisis kesan yang berlainan daripada pelbagai program perbelanjaan awam terhadap pendapatan pelbagai kumpulan isi rumah. SAM akan menjawab persoalan sama ada dan yang manakah komponen dalam perbelanjaan awam akan memberi manfaat kepada mereka yang miskin. Akaun lain seperti akaun bagi syarikat-syarikat, akaun bagi pelaburan modal swasta, dan akaun bagi transaksi dengan dunia adalah dalam bentuk agregat. Untuk menilai kesan perbelanjaan sektor awam ke atas agihan pendapatan, kebanyakan ahli ekonomi menggunakan pelbagai alat dasar polisi termasuklah model ekonometrik, analisis keberkesanan kos dan analisis kos faedah sosial.*

*Alat-alat yang sedia ada ini berasaskan kepada prosedur anggaran yang tidak mengambil kira interaksi kompleks antara kemiskinan, pengagihan pendapatan dan pembolehubah lain, serta endogeneiti agihan pendapatan. Kekuatan SAM berbanding dengan kaedah-kaedah lain yang disebut di atas bergantung kepada pembentukan model pengagihan pendapatan yang secara konsisten dan keupayaannya untuk mengesan rantaian hubungan daripada perubahan dalam permintaan kepada perubahan dalam pengeluaran, pendapatan faktor, pendapatan isi rumah dan permintaan akhir.*

## INTRODUCTION

Income distribution issues continue to attract attention in Malaysia as income inequalities between ethnic groups; states; and urban and rural areas remain wide and persistent. For more than 40 years, income inequality, represented by the Gini coefficient, was only slightly reduced to 0.441 in 2009 from 0.506 in 1970. The Gini coefficient, however, rose from 0.452 in 1999 to 0.462 in 2004. The income disparities between the Bumiputera and the Chinese; and the Bumiputera and the Indians currently remains high, with the income of the Chinese being more than one and half times greater than the income of the Bumiputera; and the income of Indians being more than one time higher than the income of the Bumiputera. The income for urban areas is more than two times greater than that of the rural areas, which has remained relatively constant since the 1970s.

This high income inequality occurs despite high public expenditures. Since independence, public expenditures, as a fiscal policy tool, have been used intensively to achieve income equality goals in Malaysia. Public expenditure grew by an average of 11.3 percent per annum for the period 1966 to 2008[1]. In the Ninth Malaysia Plan (2006-2010), the government allocated a portion of public expenditure totalling RM1,051.4 billion towards reducing income inequalities, an increase of 34.5 percent from RM781.8 billion in the Eighth Malaysia Plan (2001-2005) and RM500.8 billion during the Seventh Malaysia Plan[2]. However, the remarkable increase in public expenditure was not accompanied by a reduction in income inequality and raised the question of whether the public expenditure was allocated through appropriate channels to assist the poor.

The present paper aims to develop a social accounting matrix (SAM) framework for Malaysia to study the impact of public expenditure on income distribution in the household sector. The study will describe the structure of the SAM utilized in the present study and its detailed aggregation and disaggregation of accounts related to income inequality. In the SAM framework developed in this study, the detailed framework of the different components of public expenditure in production sectors, household sectors and other sectors is essential to analyse the differing effects of public expenditure programmes on the incomes of households in different groups.[3] The structure of the SAM presented in the present paper will answer questions concerning whether public expenditure expansion will benefit the poor and, if so, which components are most likely to be beneficial.

The application of the SAM to income distribution analysis is a method that is typically suggested by policy makers and academic economists. The continued problem of income inequality in Malaysia, as well as other developing countries, results in policy makers and academic economists amending existing macroeconomic policy tools and developing new tools in order to better understand the channels through which adjustment policies may affect the poor. The aforementioned parties believe that SAM frameworks can provide a complete picture of the impact of any adjustment policies on the system of economy, particularly the impact on income distribution. This method could provide the answer to the following questions: at present, who gets what as a result of economic activity? Who generates this income? What do the poor get from the economic activity?

## LITERATURE REVIEW

In the SAM, the distributional impact of public expenditure works through market mechanisms. As a reaction to public expenditure, optimizing firms will change their demand for factor inputs, intermediate inputs, and their supply of commodities. Changes in a firm's demand for factors will affect prices of factors of production, such as wages and non-labour income in the factor markets, and, in the end, affect household income and income distribution across households. Changes in the income of every household depend on the composition of their ownership of the factors, such as unskilled labour, skilled labour, capital and land. Changes in household income combined with changes in commodity prices will simultaneously change household expenditures on various commodities. This will affect the distribution of income and expenditure. In a general equilibrium framework, this series of mechanisms work simultaneously in inter-related markets, or, in other words, the SAM represents a detailed economy-wide circular flow of income between production activities and different institutions through the factor and product market.

[1](calculations are based on data from Bank Negara Malaysia Quarterly Bulletin (various issues))

[2](The Ninth, Eighth, and Seventh Malaysia Plans) (1996-2000)

[3]There are nine types of household groups; rural Malay, rural Chinese, rural Indian, rural others, urban Malay, urban Chinese, urban others, non-citizen.

The strength of the SAM over econometric models, cost effectiveness analyses and social cost benefit analyses lies in the modelling of the distribution of income in a consistent manner and the ability to trace chains of linkages from changes in demand to changes in production, factor incomes, household incomes, and final demands. More than any other issue, the distribution of income probably results from a complex set of relationships that require a general equilibrium model analysis, such as the SAM (Thorbecke 2000; Roland-Host and Sancho 1995; Pyatt 1991).

Nevertheless, the size and quantitative nature of the SAM model places an enormous strain on the database and limits the number of variables which can be feasibly incorporated while ensuring the accuracy of the data. In addition, the high level of disaggregation required for the model can only be assembled from censuses and surveys which are usually not conducted frequently; are conducted sporadically; and are conducted by various bodies.

The use of the SAM in income distributions gained momentum approximately thirty years ago. The surge was closely related to growing dissatisfaction with the results of growth policies in developing countries. The frustrating results of such policies, in particular with regard to their distributional impacts, shifted attention to questions concerning the processes and mechanisms that determine the relationships between the production of goods and services; income formation; and income distribution. To examine these kinds of questions, data are required that enable a comprehensive analysis of these aspects of the economic process. Only part of the data required for such analysis is provided by existing data frameworks, such as conventional national accounts and input-output tables. Initial studies of the application of SAMs to issues concerning income distribution include the study by Pyatt and Round (1977) concerning the application of a SAM to development planning generally; Adelman-Robinson (1978), who apply a SAM to income distribution in Korea; and Ahluwia and Lysy (1979), who apply a SAM to income distribution in Malaysia.

The effort to include public expenditure as a policy intervention in SAMs has received considerable attention. Keuning and Thorbecke (1989) apply a SAM to their study for the World Bank, which examines the impact of public expenditure on income distribution in Indonesia. Dorosh, Sahn and Younger (1997) include the reduction in government spending in their SAM models for the African countries of Cameroon, Gambia, Madagascar and Nigeria to simulate the effects of policy reform on the real incomes of various household groups. The study successfully demonstrates the relevance of the SAM model in highlighting and addressing issues related to public expenditure and income distribution and poverty. More recently, Agenor, Izquierdo and Fofack (2003) promote the inclusion of public expenditure in SAM models in their policy paper for the World Bank regarding developing countries, titled 'IMMPA (Integrated Macroeconomic Model for Poverty Analysis): A quantitative macroeconomic framework for the analysis of poverty reduction strategies'.

The SAM model, however, has a relatively short history in Malaysia. Among the pioneers of the SAM in Malaysia are Ramesh et al. (1980), who examine the distribution of income in the Malaysian economy in 1970. About the same time, the equilibrium model was introduced by Ahluwia and Lysy (1979). The model, however, primarily focuses upon a complex theoretical structure. The Economic Planning Unit (EPU), World Bank experts Pyatt, Round and Denes constructed a national SAM for the Malaysian economy in 1970, which distinguishes between Peninsular Malaysia (West) and the States of Sabah and Sarawak (East) in 1984. Khor (1982) applies a dual-dual approach (formal and informal sectors) in the SAM model to study income distribution and unemployment in Malaysia. More recent work was performed by Zakariah (2005), who examines theoretical issues regarding the application of SAMs in policy formulation in Malaysia.

It is important to note here that a significant limitation in all previous Malaysian SAMs, as well as many studies concerning SAMs generally, is that the research inherently focuses upon real economic activities or real accounts when studying income distribution without incorporated financial accounts in their SAM frameworks. Although Ramesh et al. (1980) incorporate the capital accounts of domestic institutions in their Malaysian SAM, capital accounts is in a consolidated form and the study does not disaggregate the capital accounts of the separate institutions. Therefore, the principal loss in information contains saving-investments or the flow of funds between institutions. Ramesh et al. (1980) state that 'to capture such detail for Malaysia, a good deal more work needs to be done and this might deserve a high priority in future developments'.

The SAM model utilized in the present study attempts to improve upon the existing Malaysian SAM by incorporating saving-investments between institutions, particularly through the disaggregation of public capital investment according to different production sectors. The inclusion of the detailed composition of public sector capital investment in the SAM framework has the potential to provide information regarding the role of the government as an intermediary among the sectors and institutions; and, thus, identify effective public sector expenditure policies for poverty reduction and income distribution. Moreover, the inclusion of the different components of public sector capital investment in the structure of the SAM attempts to strengthen SAM-based short-term forecasting models to make them more effective in tracing the implications of public expenditure adjustment policies. Furthermore, the present SAM model is based on more recent data. Data collected in 2000 provides an up-to-date coherent picture of the complexity of the Malaysian economy. The latest SAM framework applied to the Malaysian economy was produced by Pyatt et al. (1984) and based upon data from 1970. The existing SAM model has not been updated since 1984 despite the significant changes in the structure of the economy.

## STRUCTURE OF SAM TO ANALYSE THE IMPACT OF PUBLIC EXPENDITURE ON INCOME DISTRIBUTION

The construction of the SAM utilized in the present study is inspired by the works of Keuning and Thorbecke (1989); Pyatt and his associates (Pyatt 1991; Pyatt and Round 1985, 1979, 1977; Pyatt, Round and Denes 1984); Agenor et al. (2003); Emini (2002); Emini and Fofack (2004); and Kubursi (1973). The present SAM framework applies a top-down approach, which involves designing a highly aggregated SAM framework based upon available information from national accounts statistics. Then, the aggregated SAM for each account is used as a controlled value when disaggregating each of the accounts. The top-down approach is applied in the present study because of the lack of data and the approach is cost effective for the researcher.As is typical in most developing countries,lack of data is a common constraint in Malaysia because of the non-existence of data. The lack of data in the present study, however,is due to the fact that some data is not available to the public (for instance the Household Income Survey data) or data is not included in public surveys, such as inter-agent transfer payment. The use of the top-down approach is to reduce costs and save the researcher's time.

## SAM SCHEMATIC

The development of the SAM framework for Malaysia is based upon the Malaysian national account and the 2000 Input-output Table. The 2000 Input-output Table is utilized instead of the 2005 Input-output Table because the former possesses many features that are important in the construction of the SAM model, particularly the commodity-to-commodity table. A highly aggregated macro-SAM is first constructed based on the Schematic SAM. In the second stage, a disaggregated micro-SAM, called the Malaysian SAM 2000, is developed using the macro-SAM aggregates.

Following the objective of the study, twelve accounts in the Schematic SAM (Table 1) are identified. They can be grouped into five broad categories: production activities; factors of productions; institutions (household, company and public sector); indirect tax; and the rest of the world (ROW). The schematic SAM captures the inter-relationships between these accounts in the economy within a single accounting framework. The distributional income between the economic agents can be traced by looking at the flows around the Schematic SAM, which may be viewed as a systematic data system furnishing initial information on production structure; value-added payment; income distribution among agents; capital distribution; tax structure; and external flows. The Schematic SAM focuses on the production sectors, public sectors and household accounts.

The production sectors produce different sectoral goods and services for consumption by various production sectors and final users. Each production sector sells its output to other industries as an intermediate input (1,1) to the household (1,3) and the government (1,5) as the final consumer of domestic commodities; to the government (1,6) and to the private sector (1,10) as capital; and to the ROW as exports (1,11). Production sectors produce outputs by buying intermediate inputs, through the purchase of raw materials and intermediate goods and services from other sectors (1,1); employing factors of production (2,1); and importing raw materials (11,1). These accounts also pay indirect taxes to the government (7,1).

Payments received by the various types of factors of production, which are categorised into labour and capital,are distributed to various institutions in the economy, including households and companies.

In the SAM model, the household represents all people in the society. As such, the household is properly viewed as an institutional unit in the SAM. Households are often considered to be behaviourally distinct units that make economic decisions about the supply of labour and consumption expenditure. Furthermore, definitions of poverty or economic welfare are often expressed in terms of per-capita household income and consumption. The household, therefore, becomes the natural focus of the SAM analysis. Households receive factor income from production sectors in the forms of wages and other labour income (3,1); inter-household transfer (3,3); distributed profits and transfers from companies (3,4); transfers from government (3,5); and transfers from the ROW (3,11). These sources of income are utilized in the consumption of domestic commodities (1,3) and consumptions from abroad (11,3); commodity taxes (7,3) and income taxes (8,3); and inter-household transfers (3,3), with residual savings transferred to private capital accounts (10,3).

Companies are the entities that 'own' capital stock and receive profits (4,1) and non-factor income from abroad (4,11). Out of their income, they pay commodity taxes (7,4) and corporate taxes to the government (8,4); pay factorial and non-factorial income abroad (11,4); net investment abroad (12,4); and spend on distributed profits and transfer payment (3,4), with their residual savings being channelled into capital accounts (10,4).

The traditional approach assumes that public sector expenditure in the production sectors ultimately benefits the household sector. Therefore, the public sector

TABLE 1. Schematic SAM for Malaysia

| | | 1 | 2 | 3 | 4 | 5 | 6 | 7 | 8 | 9 | 10 | 11 | 12 | |
|---|---|---|---|---|---|---|---|---|---|---|---|---|---|---|
| | | Production sectors | Factor of production | Household | Companies | Public current expenditure on domestic commodities and households | Public capital investments in production activities | Indirect taxes | Public current | Public capital | Private capital | ROW current | ROW capital | Total |
| 1 | Production sectors | Raw materials purchase of domestic commodities (IntG) | | Household consumption of domestic commodities (HCd) | | Public expenditure on domestic commodities (PCd) | Public investment in production activities (PId) | | | | Private investment in production activities (PrId) | Exports (X) | | Gross output (Y) |
| 2 | Factor of production | Value added (Va) | | | | | | | | | | | | Total factor of production payment (TFPy) |
| 3 | Households | | Factor income to households (FincH) | Inter household transfers (Tr) | Distributed profit and current transfer (Div) | Transfer (Tr) | | | | | | Social benefits received from abroad (YRow) | | Total household income (YH) |
| 4 | Companies | | Business corporate profits (CorProf) | | | Transfer (Tr) | | | | | | Non-factor income from abroad (NfYRow) | | Total company income (YCor) |
| 5 | Public current expenditure on domestic commodities and households | | | | | | | | Public current expenditure (PCd) | | | | | Total public current expenditure (TPCd) |
| 6 | Public capital investment in production activities | | | | | | | | | Public investments (PId) | | | | Total public capital investments (TPId) |

cont.

continue

| | | | | | | | | | | | | | | |
|---|---|---|---|---|---|---|---|---|---|---|---|---|---|---|
| 7 | Indirect tax | Commodity taxes (TxGd) | | Commodity taxes (TxGh) | Commodity taxes (TxGcor) | | | | | Capital taxes (TxGPCap) | Capital taxes (TxGPrCap) | Exports/ imports levy (LV) | | Total indirect taxes (TxG) |
| 8 | Public current | | | Income taxes (IncTax) | Corporate tax (TxCor) | | | indirect taxes (TxG) | | | | Non-factor income from abroad (NfYRow) | | Total public current (TPC) |
| 9 | Public capital | | | | | | | | Public current surplus (Surp) | | Domestic capital (CapDom) | | Capital transfer from abroad (CapPRw) | Public capital (TPK) |
| 10 | Private capital | | | Household saving (SH) | Companies saving (SCor) | | | | | | | | Capital transfer from abroad (CapPrRw) | Private capital (TPrK) |
| 11 | ROW current | Import of Raw materials (Mr) | | Household consumption of imported commodities (Mh) | | | | | Public consumption of imported commodities and current transfer (MP) | Public investment on imported capital (MCapP) | Private investment on imported capital (MCapPr) | Balance of goods and services (BoGS) | | Total current ROW (TCRow) |
| 12 | ROW capital | | | | Net investment abroad (InRow) | | | | | | | Balance of payment (BoP) | | Total ROW capital (TCapRow) |
| | Total | Gross inputs (Y) | Total factor payments (TFPy) | Total household expenditure (Yh) | Total company expenditure (YCor) | Total public current expenditure on domestic commodities and households (TPCd) | Total public capital investments on domestic commodities (TPId) | Indirect taxes (TxG) | Total public current (TPC) | Total public capital (TPK) | Total private capital (TPrK) | Total ROW current (TCRow) | Total ROW capital (TCapRw) | |

TABLE 2. Malaysia Macro-SAM

| | | 1 | 2 | 3 | 4 | 5 | 6 | 7 | 8 | 9 | 10 | 11 | 12 | 13 | |
|---|---|---|---|---|---|---|---|---|---|---|---|---|---|---|---|
| | (RM Billion) | Production sectors | Factor of production | Households | Companies | Public current expenditure | Public capital investment | Indirect taxes | Public current | Public capital | Private capital | Changes in inven-tory | ROW current | ROW capital | Total |
| 1 | Production sectors | 271.7 | | 116.583 | | 34.862 | 10.597 | | | | 40.784 | 22.92 | 399.379 | | 896.828 |
| 2 | Factor of production | 345.27 | | | | | | | | | | | | | 345.27 |
| 3 | Household | | 152.303 | 3.819 | 22.734 | 7.861 | | | | | | | 1..301 | | 188.018 |
| 4 | Companies | | 192.967 | | | | | | | | | | 8.674 | | 201.641 |
| 5 | Public current expenditure | | | | | | | | 42.723 | | | | | | 42.723 |
| 6 | Public capital investment | | | | | | | | | 10.597 | | | | | 10.597 |
| 7 | Indirect tax | 8.407 | | 9.213 | 0..3 | | | | | 0.303 | 1.168 | | 1.086 | | 20.477 |
| 8 | Public current | | | 7.015 | 27.263 | | | 20.477 | | | | | 0.444 | | 55.199 |
| 9 | Public capital | | | | | | | | 11.557 | | 10.897 | | | 0.864 | 23..32 |
| 10 | Private capital | | | 28.784 | 72.593 | | | | | | | | | -0.881 | 100.64 |
| 11 | Changes in inventory | | | | | | | | | 4.73 | 18.19 | | | | 22.92 |
| 12 | ROW current | 271.451 | | 22.605 | 46.494 | | | | 0.772 | 7.69 | 29.598 | | 26.559 | | 427.024 |
| 13 | ROW capital | | | | 32.257 | | | | | | | | -32.274 | | -0.017 |
| | Total | 896.828 | 345.27 | 188.019 | 201.642 | 42.723 | 10.597 | 20.477 | 55.199 | 23.32 | 100.64 | 22.92 | 405.169 | -0.017 | 1967.52 |

*Source:* Input-output Table 2000, 2005, DOS; Final National Account Statistics Malaysia 2000,2005, DOS; Distribution and Use of Income Account and Capital Account 2000-2002, 2007, DOS; Economic Reports 2000/2001, Treasury Malaysia; BNM Quarterly Bulletin, Fourth Quarter 2001, Ministry of Finance.

Note: Table based on the Schematic SAM.

plays an important role in the redistribution process. The public sector account consists of two distinct accounts: current accounts and capital accounts. The public current account receives direct taxes from the household (8,3) and companies (8,4); indirect taxes (8,7); and non-factor income from abroad (8,11). The public current account allocates its current expenditures to the purchase of products and services provided by the domestic production activities account (1,5); transfers to households (3,5); and the import of commodities (11,8). The remaining savings or current account surplus is transferred to the public capital account (9,8). Together with the domestic (9,10) and foreign (9,12) funds, the capital account allocates its investments to domestic production activities (1,6), capital taxes (7,9) and imported capital goods from the ROW (11,9).

The private sector receives capital from household savings (10,3), company savings (10,4), and from external sources (10,12). Private capital is then allocated to investments in production activities (1,10); capital taxes (7,10); as a source of domestic capital for the public sector (9,10); and the import of capital goods (11,10).

Finally, the transactions between domestic and the foreign residents are recorded in the ROW account. Malaysian households, companies and government consumption expenditures on imported final goods and imported capital goods contributed to the ROW receipt. The Malaysian economy receives income from the ROW in the form of exports; factor and non-factor income earned; and export and import levies. When focusing on the distribution of income, it is important to keep track of the distinct flow of income between the different institutions and the ROW.

Based on this schematic SAM, a macro-aggregate Malaysian SAM is designed, as shown in Table 2. The Malaysian macro-aggregate SAM shows an aggregate value of each account that can act as a control value for designing a detailed Malaysian SAM. The level of disaggregation of the individual account depends primarily upon the specific question that the SAM is expected to answer. In this case, studying the impact of public expenditure on income distribution necessitates an examination of the composition of public expenditure; different categories of production activities; household inter-ethnic disparity; and urban-rural bias.Together, these elements capture the different dimensions of income inequality. Disaggregation captures how changes in public expenditure that affect various production structures are transmitted to the household sector.

As the treatment of public expenditure is of particular interest, the framework incorporates a detailed breakdown of public expenditure in the various sectors. An essential point in this context is not only that a larger aggregate public expenditure today might generate a higher aggregate output tomorrow, but that the pattern and destination of public expenditure may be more important determinants of future equity and growth.

The account of various production sectors is crucial in the present SAM model because public expenditure affects income distribution through the transmission of investment and consumption of the household sector by investing in or purchasing from various production sectors.The companies in a sector that receive higher public spending will earn higher profits from increases in their sales than those companies in sectors which receive low profit spending.Consequently, stockholders and workers in those sectors, respectively, *ceteris paribus*, will receive higher dividends and higher wages.

In relation to the income distribution analysis, the present framework emphasizes the household group differentiation in the transmission of public expenditure to impoverished households. The disaggregation of the household sector can capture the changes in the production structures due to public expenditure transmission to the household sector. The disaggregation of the household is based on socio-economic groups, rather than on income levels. Being a multi-racial country, it is crucial to distinguish between four major ethnic groups for the household: Malays, Chinese, Indians and others.[4] This disaggregation is very important as income equality among the ethnic groups has been an important government development strategy since independence. Additionally, due to the fact that the majority of impoverished households are located in rural areas, the distinction between households in rural and urban areas is also very important. The urban-rural area disaggregation is useful because the distinction captures many aspects of duality. Typically, the urban sector contains a labour force that has substantial skill specificity, good working conditions, high pay and high job security. The rural sector, on the other hand, consists of a labour force that possesses little skill specificity, poor working conditions, low pay and little job security. The distinction is also made between citizens and non-citizens, since it is believed that the number of foreign workers has influenced significantly the patterns of the domestic labour force due to the fact that most foreign workers are employed in the plantation and agriculture sectors; and in construction activities.

A detailed framework of Malaysia's SAM 2000 is designed, producing the 51 x 51 matrix accounts that disaggregate the macro-aggregate SAM (table in the appendix). The most notable feature of the Malaysian SAM is the incorporation of detailed disaggregation of public capital investments, which demonstrates the relationships between the public sector financial account and the real account. The disaggregation shows how the use of public funds in the different segments in the economy can generate different levels of income for different household groups. For instance, the public capital investment in agriculture and rural development causes a demand in the agricultural sector for labour. This, in turn, results in demand for wages based on their factor ownership and the generation of income or output in that sector.

[4]The 'Others' group consists of minorities that are principally located in East Malaysia, such as the Iban, Kadazan, Bajau, and Murut.

Generally, the principal characteristics and features of the present SAM that distinguish it from the existing SAM are as follows:

1. In the SAM utilized in the present study, public expenditures consist of a) seven categories of public current expenditure; and b) seven categories of public investment, including disaggregation in the public capital account. This disaggregation into various categories is important as it enables the examination of different effects of public expenditure on household incomes distributions.In the existing SAM, public expenditure is not divided into categories. It is consolidated in one account.
2. In the SAM utilized in the present study households consist of nine socio-economic household groups distinguished according to whether the household is located in rural or urban areas; the ethnicity of the household; and whether the household is comprised of citizens and non-citizens. Other institutions are the public sector and the companies. The SAM therefore contains the detailed breakdown of the household groups which are categorised according to the various ethnic groups that reside in the respective urban and rural areas and also includes the non-citizens. In the existing SAM, the household groups are categorised according to the ethnic groups and the urban and rural areas. For example, Ramesh et al. (1980) disaggregates households into 6 groups: Malay, Chinese, Indians, others, rural and urban. This disaggregation is sufficient as long as the analysis is simply to compare and contrast the impact of public expenditure between these socioeconomic classes. However, such a disaggregation prevents more precise questions from being answered, such as how different is the impact of public expenditure between the poor and the rich; how much will inequality change; and how much is poverty incidence changed by the policies.
3. In the SAM utilized in the present study, other accounts include 18 production activities,private capital, ROW capital and ROW current. Disaggregation of the capital account is important to see the separate effects of the public and private capitals on household income distribution. Similarly, the disaggregation of the ROW account will enable the examination of the effects of ROW capital and ROW current on household income distribution separately. Meanwhile, in the existing SAM, the private capital account is lumped together with the public capital account, forming a single account; and the ROW account is in the form of a single account.

## CONCLUDING REMARKS

Public sector expenditure programmes represent major efforts by the government to reduce wide economic and social imbalances among ethnic groups and regions. The purpose of this paper is to construct a Malaysian SAM to study the income distribution impact of public sector expenditure. The detailed structure of the SAM utilized in the present study allows for the analysis of the impact of the public sector expenditure expansion on the economy; and inter-sectoral and inter-institutional income linkages, particularly among poor households.

A framework that incorporates different classes or components of public expenditure can demonstrate the differing effects of different public sector expenditure programmes on various economic variables in the system as these components absorb different sector purchases and, therefore, exhibit differences in income generation and distribution of income. Additionally, a detailed framework of different sectors that characterizes various production sectors of the economy which could exhibit differences in income generation; and a detailed framework of different household groups that characterizes the income inequalities between ethnic groups and regions which could exhibit a distribution of income are required to complement the framework.

Organizing public sector expenditure by various classes in the SAM utilized in the present study represents an improvement over the extant literature on this subject as it emphasizes the importance of the various components of public sector expenditure. More obviously, the incorporation of government capital investment in the SAM can reflect the role of the government in the income distribution through its role as a stimulus of demand in the economy.

Mukaramah-Harun
Universiti Utara Malaysia
mukaramah@uum.edu.my

A.R. Zakariah
Malaysian Institute of Economic Research
zakariah@mier.po.my

M. Azali
Universiti Putra Malaysia
azali@putra.upm.edu.my

APPENDIX

| | RM Million | 1<br>Agricul & livestock | 2<br>Forestry | 3<br>Fish | 4<br>Mining & quarrying | 5<br>Manufac | 6<br>Electricity Gas & Water |
|---|---|---|---|---|---|---|---|
| 1 | Agriculture and livestock | 2,519.43 | 0.00 | 4.12 | 0.00 | 8,552.23 | 0.20 |
| 2 | Forestry and logging products | 0.03 | 300.56 | 0.00 | 0.01 | 7,047.88 | 0.00 |
| 3 | Fish etc. | 0.00 | 0.00 | 510.89 | 0.00 | 868.01 | 0.00 |
| 4 | Mining & quarrying | 4.25 | 0.00 | 0.00 | 522.91 | 12,989.63 | 0.01 |
| 5 | Manufacturing | 3,634.70 | 877.59 | 1,258.25 | 1,096.51 | 91,409.53 | 1,938.18 |
| 6 | Electricity, gas & water | 97.91 | 0.00 | 46.67 | 84.87 | 8,059.28 | 908.52 |
| 7 | Building and construcions | 61.68 | 0.00 | 12.09 | 97.57 | 1,400.37 | 90.54 |
| 8 | Wholesale and retail trade | 668.12 | 346.95 | 271.00 | 322.11 | 30,168.63 | 405.00 |
| 9 | Hotel& restaurant | 38.14 | 0.00 | 0.47 | 84.76 | 883.25 | 81.59 |
| 10 | Transport and communication | 254.52 | 35.35 | 31.79 | 591.91 | 5,364.39 | 132.57 |
| 11 | Financial, insurance and real estate | 44.84 | 0.00 | 5.71 | 141.17 | 518.74 | 243.72 |
| 12 | Business services | 81.83 | 33.67 | 8.83 | 506.09 | 1,976.02 | 1,111.96 |
| 13 | Education | 0.00 | 0.00 | 0.00 | 13.42 | 0.16 | 39.60 |
| 14 | Health | 14.18 | 0.00 | 0.00 | 0.00 | 0.10 | 0.00 |
| 15 | Other private services | 20.90 | 42.62 | 10.77 | 37.78 | 185.02 | 25.50 |
| 16 | General administration | 0.45 | 0.00 | 0.00 | 0.00 | 0.56 | 0.15 |
| 17 | Public order and defence | 0.00 | 0.00 | 0.00 | 0.00 | 0.00 | 0.00 |
| 18 | Other Public Administration | 4.39 | 0.07 | 0.00 | 0.00 | 4.27 | 0.00 |
| 19 | Factor of production | 11,820.55 | 8,252.49 | 3,096.42 | 35,843.36 | 107,320.35 | 10,767.51 |
| 20 | Rural Malay | 0.00 | 0.00 | 0.00 | 0.00 | 0.00 | 0.00 |
| 21 | Rural Chinese | 0.00 | 0.00 | 0.00 | 0.00 | 0.00 | 0.00 |
| 22 | Rural Indian | 0.00 | 0.00 | 0.00 | 0.00 | 0.00 | 0.00 |
| 23 | Rural Others | 0.00 | 0.00 | 0.00 | 0.00 | 0.00 | 0.00 |
| 24 | Urban Malay | 0.00 | 0.00 | 0.00 | 0.00 | 0.00 | 0.00 |
| 25 | Urban Chinese | 0.00 | 0.00 | 0.00 | 0.00 | 0.00 | 0.00 |
| 26 | Urban Indian | 0.00 | 0.00 | 0.00 | 0.00 | 0.00 | 0.00 |
| 27 | Urban others | 0.00 | 0.00 | 0.00 | 0.00 | 0.00 | 0.00 |
| 28 | Non-citizen | 0.00 | 0.00 | 0.00 | 0.00 | 0.00 | 0.00 |
| 29 | Companies | 0.00 | 0.00 | 0.00 | 0.00 | 0.00 | 0.00 |
| 30 | Pub Exp Agriculture | 0.00 | 0.00 | 0.00 | 0.00 | 0.00 | 0.00 |
| 31 | Pub Exp Education | 0.00 | 0.00 | 0.00 | 0.00 | 0.00 | 0.00 |
| 32 | Pub Exp Health | 0.00 | 0.00 | 0.00 | 0.00 | 0.00 | 0.00 |
| 33 | Public Exp Administration | 0.00 | 0.00 | 0.00 | 0.00 | 0.00 | 0.00 |
| 34 | Public Exp Pub Order & Defense | 0.00 | 0.00 | 0.00 | 0.00 | 0.00 | 0.00 |
| 35 | Pun Exp Other Public Admin | 0.00 | 0.00 | 0.00 | 0.00 | 0.00 | 0.00 |
| 36 | Pub Exp Household Tran | 0.00 | 0.00 | 0.00 | 0.00 | 0.00 | 0.00 |
| 37 | Pub Inv Agric & Rural Development | 0.00 | 0.00 | 0.00 | 0.00 | 0.00 | 0.00 |
| 38 | Pub Inv Industry | 0.00 | 0.00 | 0.00 | 0.00 | 0.00 | 0.00 |
| 39 | Pub Inv Trade | 0.00 | 0.00 | 0.00 | 0.00 | 0.00 | 0.00 |
| 40 | Pub Inv Transportation & Communication | 0.00 | 0.00 | 0.00 | 0.00 | 0.00 | 0.00 |
| 41 | Pub Inv Educ & health | 0.00 | 0.00 | 0.00 | 0.00 | 0.00 | 0.00 |
| 42 | Pub Inv Administration | 0.00 | 0.00 | 0.00 | 0.00 | 0.00 | 0.00 |
| 43 | PubInvOthers | 0.00 | 0.00 | 0.00 | 0.00 | 0.00 | 0.00 |
| 44 | Commodities Taxes (Domestic) | 38.97 | 6.55 | 6.43 | 32.10 | 1,993.46 | 65.99 |
| 45 | Commodities Taxes (Imports) | 19.83 | 15.59 | 7.44 | 18.64 | 2,590.41 | 33.38 |
| 46 | Public Current | 0.00 | 0.00 | 0.00 | 0.00 | 0.00 | 0.00 |
| 47 | Public Capital | 0.00 | 0.00 | 0.00 | 0.00 | 0.00 | 0.00 |
| 48 | Private Capital | 0.00 | 0.00 | 0.00 | 0.00 | 0.00 | 0.00 |
| 49 | Changes in inventories | 0.00 | 0.00 | 0.00 | 0.00 | 0.00 | 0.00 |
| 50 | ROW current | 1,583.33 | 1,279.25 | 181.93 | 4,464.69 | 213,974.03 | 1,535.68 |
| 51 | ROW capital | 0.00 | 0.00 | 0.00 | 0.00 | 0.00 | 0.00 |
| | TOTAL | 20,908.05 | 11,190.69 | 5,452.83 | 43,857.89 | 495,306.31 | 17,380.11 |

Continue

| | | 7 | 8 | 9 | 10 | 11 | 12 |
|---|---|---|---|---|---|---|---|
| | | Building & Constr | Wholesale & retail Tra | Hotel & Restaur | Transport & communicat | Financ, Insuran & Real Estate | Business Services |
| 1 | Agriculture and livestock | 12.06 | 11.99 | 1,109.07 | 11.53 | 11.74 | 27.31 |
| 2 | Forestry and logging products | 18.63 | 15.33 | 0.00 | 0.01 | 9.85 | 14.23 |
| 3 | Fish etc. | 2.98 | 4.94 | 971.85 | 5.39 | 6.34 | 10.49 |
| 4 | Mining & quarrying | 1,435.50 | 4.84 | 0.00 | 0.26 | 6.16 | 41.30 |
| 5 | Manufacturing | 12,629.37 | 964.60 | 4,012.70 | 5,855.96 | 1,024.08 | 1,279.08 |
| 6 | Electricity, gas & water | 191.64 | 828.08 | 827.08 | 536.60 | 816.26 | 188.52 |
| 7 | Building and construcions | 195.52 | 518.82 | 147.85 | 116.65 | 266.10 | 87.92 |
| 8 | Wholesale and retail trade | 2,386.92 | 235.97 | 672.61 | 1,055.39 | 222.16 | 472.86 |
| 9 | Hotel& restaurant | 235.52 | 1,802.63 | 319.10 | 1,978.67 | 278.04 | 331.05 |
| 10 | Transport and communication | 1,011.35 | 1,157.45 | 626.94 | 5,021.66 | 2,003.91 | 891.18 |
| 11 | Financial, insurance and real estate | 1,007.45 | 2,133.46 | 182.22 | 1,011.93 | 4,963.02 | 597.18 |
| 12 | Business services | 585.47 | 644.54 | 311.05 | 1,130.61 | 873.37 | 929.82 |
| 13 | Education | 9.91 | 0.04 | 0.00 | 0.20 | 1.76 | 5.04 |
| 14 | Health | 0.00 | 0.17 | 0.01 | 0.29 | 0.04 | 0.02 |
| 15 | Other private services | 62.20 | 152.20 | 103.26 | 411.80 | 75.60 | 121.60 |
| 16 | General administration | 0.03 | 0.02 | 0.00 | 0.14 | 0.46 | 7.51 |
| 17 | Public order and defence | 0.00 | 0.00 | 0.00 | 0.00 | 0.00 | 0.00 |
| 18 | Other Public Administration | 0.01 | 0.41 | 0.00 | 1.86 | 1.13 | 0.05 |
| 19 | Factor of production | 14,272.56 | 38,145.48 | 7,420.30 | 22,178.12 | 45,883.36 | 11,274.85 |
| 20 | Rural Malay | 0.00 | 0.00 | 0.00 | 0.00 | 0.00 | 0.00 |
| 21 | Rural Chinese | 0.00 | 0.00 | 0.00 | 0.00 | 0.00 | 0.00 |
| 22 | Rural Indian | 0.00 | 0.00 | 0.00 | 0.00 | 0.00 | 0.00 |
| 23 | Rural Others | 0.00 | 0.00 | 0.00 | 0.00 | 0.00 | 0.00 |
| 24 | Urban Malay | 0.00 | 0.00 | 0.00 | 0.00 | 0.00 | 0.00 |
| 25 | Urban Chinese | 0.00 | 0.00 | 0.00 | 0.00 | 0.00 | 0.00 |
| 26 | Urban Indian | 0.00 | 0.00 | 0.00 | 0.00 | 0.00 | 0.00 |
| 27 | Urban others | 0.00 | 0.00 | 0.00 | 0.00 | 0.00 | 0.00 |
| 28 | Non-citizen | 0.00 | 0.00 | 0.00 | 0.00 | 0.00 | 0.00 |
| 29 | Companies | 0.00 | 0.00 | 0.00 | 0.00 | 0.00 | 0.00 |
| 30 | PubExpAgriculture | 0.00 | 0.00 | 0.00 | 0.00 | 0.00 | 0.00 |
| 31 | PubExpEducation | 0.00 | 0.00 | 0.00 | 0.00 | 0.00 | 0.00 |
| 32 | PubExpHealth | 0.00 | 0.00 | 0.00 | 0.00 | 0.00 | 0.00 |
| 33 | PublicExpAdministration | 0.00 | 0.00 | 0.00 | 0.00 | 0.00 | 0.00 |
| 34 | PublicExpPubOrder&Defense | 0.00 | 0.00 | 0.00 | 0.00 | 0.00 | 0.00 |
| 35 | PunExpOtherPublicAdmin | 0.00 | 0.00 | 0.00 | 0.00 | 0.00 | 0.00 |
| 36 | PubExpHouseholdTran | 0.00 | 0.00 | 0.00 | 0.00 | 0.00 | 0.00 |
| 37 | PubInvAgric&RuralDevelopment | 0.00 | 0.00 | 0.00 | 0.00 | 0.00 | 0.00 |
| 38 | PubInvIndustry | 0.00 | 0.00 | 0.00 | 0.00 | 0.00 | 0.00 |
| 39 | PubInvTrade | 0.00 | 0.00 | 0.00 | 0.00 | 0.00 | 0.00 |
| 40 | PubInvTransportation& Communication | 0.00 | 0.00 | 0.00 | 0.00 | 0.00 | 0.00 |
| 41 | PubInvEduc&health | 0.00 | 0.00 | 0.00 | 0.00 | 0.00 | 0.00 |
| 42 | PubInvAdministration | 0.00 | 0.00 | 0.00 | 0.00 | 0.00 | 0.00 |
| 43 | PubInvOthers | 0.00 | 0.00 | 0.00 | 0.00 | 0.00 | 0.00 |
| 44 | Commodities Taxes (Domestic) | 349.18 | 422.30 | 131.07 | 444.73 | 1,140.22 | 132.50 |
| 45 | Commodities Taxes (Imports) | 94.07 | 13.61 | 51.22 | 196.99 | 7.28 | 32.92 |
| 46 | Public Current | 0.00 | 0.00 | 0.00 | 0.00 | 0.00 | 0.00 |
| 47 | Public Capital | 0.00 | 0.00 | 0.00 | 0.00 | 0.00 | 0.00 |
| 48 | Private Capital | 0.00 | 0.00 | 0.00 | 0.00 | 0.00 | 0.00 |
| 49 | Changes in inventories | 0.00 | 0.00 | 0.00 | 0.00 | 0.00 | 0.00 |
| 50 | ROW current | 10,590.67 | 5,290.72 | 4,018.36 | 12,084.94 | 3,299.89 | 5,256.03 |
| 51 | ROW capital | 0.00 | 0.00 | 0.00 | 0.00 | 0.00 | 0.00 |
| | TOTAL | 45,091.03 | 52,347.60 | 20,904.69 | 52,043.73 | 60,890.76 | 21,701.46 |

Continue

| | | 13<br>Eductn | 14<br>Health | 15<br>Other Priv. Serv | 16<br>General administr | 17<br>Public Order & Defense | 18<br>Other public admin |
|---|---|---|---|---|---|---|---|
| 1 | Agriculture and livestock | 1.31 | 4.08 | 122.00 | 0.14 | 0.10 | 0.22 |
| 2 | Forestry and logging products | 0.13 | 0.06 | 46.10 | 0.02 | 0.01 | 0.17 |
| 3 | Fish etc. | 0.00 | 1.95 | 81.40 | 0.00 | 0.00 | 0.00 |
| 4 | Mining & quarrying | 0.13 | 0.04 | 40.70 | 0.07 | 0.27 | 0.40 |
| 5 | Manufacturing | 651.58 | 383.00 | 1,001.60 | 354.86 | 1,105.52 | 230.11 |
| 6 | Electricity, gas & water | 219.31 | 183.04 | 220.63 | 149.86 | 147.79 | 103.17 |
| 7 | Building and construcions | 26.02 | 20.58 | 188.99 | 45.89 | 78.45 | 62.15 |
| 8 | Wholesale and retail trade | 82.31 | 248.70 | 331.98 | 62.13 | 123.29 | 18.21 |
| 9 | Hotel& restaurant | 133.87 | 41.70 | 129.05 | 175.82 | 117.62 | 72.26 |
| 10 | Transport and communication | 165.57 | 170.19 | 287.43 | 242.39 | 201.79 | 130.92 |
| 11 | Financial, insurance and real estate | 151.90 | 80.38 | 275.49 | 297.78 | 44.27 | 139.35 |
| 12 | Business services | 318.57 | 192.05 | 333.71 | 370.64 | 128.18 | 415.98 |
| 13 | Education | 0.31 | 0.00 | 0.25 | 0.00 | 0.00 | 0.00 |
| 14 | Health | 0.00 | 4.14 | 1.03 | 0.00 | 0.00 | 0.03 |
| 15 | Other private services | 75.47 | 78.50 | 511.10 | 85.67 | 75.23 | 89.92 |
| 16 | General administration | 0.00 | 0.03 | 20.64 | 0.13 | 282.86 | 87.43 |
| 17 | Public order and defence | 0.00 | 0.00 | 0.00 | 0.00 | 36.46 | 0.00 |
| 18 | Other Public Administration | 0.00 | 0.00 | 0.02 | 0.00 | 0.00 | 0.00 |
| 19 | Factor of production | 10,563.16 | 3,779.19 | 5,049.14 | 2,818.35 | 4,456.43 | 2,328.50 |
| 20 | Rural Malay | 0.00 | 0.00 | 0.00 | 0.00 | 0.00 | 0.00 |
| 21 | Rural Chinese | 0.00 | 0.00 | 0.00 | 0.00 | 0.00 | 0.00 |
| 22 | Rural Indian | 0.00 | 0.00 | 0.00 | 0.00 | 0.00 | 0.00 |
| 23 | Rural Others | 0.00 | 0.00 | 0.00 | 0.00 | 0.00 | 0.00 |
| 24 | Urban Malay | 0.00 | 0.00 | 0.00 | 0.00 | 0.00 | 0.00 |
| 25 | Urban Chinese | 0.00 | 0.00 | 0.00 | 0.00 | 0.00 | 0.00 |
| 26 | Urban Indian | 0.00 | 0.00 | 0.00 | 0.00 | 0.00 | 0.00 |
| 27 | Urban others | 0.00 | 0.00 | 0.00 | 0.00 | 0.00 | 0.00 |
| 28 | Non-citizen | 0.00 | 0.00 | 0.00 | 0.00 | 0.00 | 0.00 |
| 29 | Companies | 0.00 | 0.00 | 0.00 | 0.00 | 0.00 | 0.00 |
| 30 | PubExpAgriculture | 0.00 | 0.00 | 0.00 | 0.00 | 0.00 | 0.00 |
| 31 | PubExpEducation | 0.00 | 0.00 | 0.00 | 0.00 | 0.00 | 0.00 |
| 32 | PubExpHealth | 0.00 | 0.00 | 0.00 | 0.00 | 0.00 | 0.00 |
| 33 | PublicExpAdministration | 0.00 | 0.00 | 0.00 | 0.00 | 0.00 | 0.00 |
| 34 | PublicExpPubOrder&Defense | 0.00 | 0.00 | 0.00 | 0.00 | 0.00 | 0.00 |
| 35 | PunExpOtherPublicAdmin | 0.00 | 0.00 | 0.00 | 0.00 | 0.00 | 0.00 |
| 36 | PubExpHouseholdTran | 0.00 | 0.00 | 0.00 | 0.00 | 0.00 | 0.00 |
| 37 | PubInvAgric&RuralDevelopment | 0.00 | 0.00 | 0.00 | 0.00 | 0.00 | 0.00 |
| 38 | PubInvIndustry | 0.00 | 0.00 | 0.00 | 0.00 | 0.00 | 0.00 |
| 39 | PubInvTrade | 0.00 | 0.00 | 0.00 | 0.00 | 0.00 | 0.00 |
| 40 | PubInvTransportation& Communication | 0.00 | 0.00 | 0.00 | 0.00 | 0.00 | 0.00 |
| 41 | PubInvEduc&health | 0.00 | 0.00 | 0.00 | 0.00 | 0.00 | 0.00 |
| 42 | PubInvAdministration | 0.00 | 0.00 | 0.00 | 0.00 | 0.00 | 0.00 |
| 43 | PubInvOthers | 0.00 | 0.00 | 0.00 | 0.00 | 0.00 | 0.00 |
| 44 | Commodities Taxes (Domestic) | 45.69 | 42.82 | 150.80 | 104.13 | 39.84 | 85.74 |
| 45 | Commodities Taxes (Imports) | 9.53 | 5.56 | 51.50 | 4.80 | 17.44 | 3.94 |
| 46 | Public Current | 0.00 | 0.00 | 0.00 | 0.00 | 0.00 | 0.00 |
| 47 | Public Capital | 0.00 | 0.00 | 0.00 | 0.00 | 0.00 | 0.00 |
| 48 | Private Capital | 0.00 | 0.00 | 0.00 | 0.00 | 0.00 | 0.00 |
| 49 | Changes in inventories | 0.00 | 0.00 | 0.00 | 0.00 | 0.00 | 0.00 |
| 50 | ROW current | 1,024.13 | 2,080.64 | 1,613.40 | 743.45 | 1,685.17 | 744.62 |
| 51 | ROW capital | 0.00 | 0.00 | 0.00 | 0.00 | 0.00 | 0.00 |
| | TOTAL | 13,468.99 | 7,316.64 | 10,456.95 | 5,456.14 | 8,540.72 | 4,513.14 |

Continue

| | | 19 | 20 | 21 | 22 | 23 | 24 | 25 |
|---|---|---|---|---|---|---|---|---|
| | | Factor of production | Rural Malay expenditure | Rural Chinese expenditure | Rural Indian expenditure | Rural others expenditure | Urban Malay expenditure | Urban Chinese expenditure |
| 1 | Agriculture and livestock | 0.00 | 1,677.83 | 165.96 | 152.72 | 12.11 | 746.55 | 1,286.20 |
| 2 | Forestry and logging products | 0.00 | 0.00 | 0.00 | 0.00 | 0.00 | 0.00 | 0.00 |
| 3 | Fish etc. | 0.00 | 643.60 | 79.07 | 7.53 | 1.94 | 479.70 | 255.14 |
| 4 | Mining & quarrying | 0.00 | 0.00 | 0.00 | 0.00 | 0.00 | 0.00 | 0.00 |
| 5 | Manufacturing | 0.00 | 8,521.65 | 1,046.13 | 399.28 | 17.39 | 9,557.48 | 13,755.07 |
| 6 | Electricity, gas & water | 0.00 | 823.04 | 19.03 | 29.67 | 3.70 | 1,812.05 | 508.27 |
| 7 | Building and construcions | 0.00 | 62.28 | 13.45 | 1.39 | 0.24 | 69.86 | 194.24 |
| 8 | Wholesale and retail trade | 0.00 | 208.39 | 61.80 | 5.65 | 0.69 | 365.34 | 1,297.94 |
| 9 | Hotel& restaurant | 0.00 | 2,482.36 | 301.09 | 56.89 | 9.99 | 4,211.36 | 5,591.72 |
| 10 | Transport and communication | 0.00 | 2,744.14 | 333.77 | 144.66 | 10.02 | 4,960.27 | 3,385.26 |
| 11 | Financial, insurance and real estate | 0.00 | 2,154.11 | 325.49 | 51.87 | 6.09 | 8,459.30 | 11,194.41 |
| 12 | Business services | 0.00 | 94.69 | 13.07 | 3.91 | 0.30 | 212.64 | 366.18 |
| 13 | Education | 0.00 | 643.22 | 21.04 | 6.30 | 2.84 | 749.22 | 332.80 |
| 14 | Health | 0.00 | 540.22 | 20.66 | 12.93 | 2.61 | 1,142.22 | 825.19 |
| 15 | Other private services | 0.00 | 763.29 | 196.84 | 22.40 | 5.23 | 1,288.35 | 3,200.16 |
| 16 | General administration | 0.00 | 0.00 | 0.00 | 0.00 | 0.00 | 0.00 | 0.00 |
| 17 | Public order and defence | 0.00 | 0.00 | 0.00 | 0.00 | 0.00 | 0.00 | 0.00 |
| 18 | Other Public Administration | 0.00 | 0.00 | 0.00 | 0.00 | 0.00 | 0.00 | 0.00 |
| 19 | Factor of production | 0.00 | 0.00 | 0.00 | 0.00 | 0.00 | 0.00 | 0.00 |
| 20 | Rural Malay | 28,354.49 | 368.12 | 0.00 | 0.00 | 0.00 | 1,071.28 | 0.00 |
| 21 | Rural Chinese | 6,636.75 | 0.00 | 37.01 | 0.00 | 0.00 | 0.00 | 117.89 |
| 22 | Rural Indian | 2,008.67 | 0.00 | 0.00 | 304.18 | 0.00 | 0.00 | 0.00 |
| 23 | Rural Others | 241.24 | 0.00 | 0.00 | 0.00 | 71.96 | 0.00 | 0.00 |
| 24 | Urban Malay | 42,905.71 | 368.12 | 0.00 | 0.00 | 0.00 | 126.49 | 0.00 |
| 25 | Urban Chinese | 55,142.39 | 0.00 | 37.01 | 0.00 | 0.00 | 0.00 | 11.62 |
| 26 | Urban Indian | 10,880.40 | 0.00 | 0.00 | 304.18 | 0.00 | 0.00 | 0.00 |
| 27 | Urban others | 415.41 | 0.00 | 0.00 | 0.00 | 71.96 | 0.00 | 0.00 |
| 28 | Non-citizen | 5,717.58 | 0.00 | 0.00 | 0.00 | 0.00 | 0.00 | 0.00 |
| 29 | Companies | 192,967.47 | 0.00 | 0.00 | 0.00 | 0.00 | 0.00 | 0.00 |
| 30 | PubExpAgriculture | 0.00 | 0.00 | 0.00 | 0.00 | 0.00 | 0.00 | 0.00 |
| 31 | PubExpEducation | 0.00 | 0.00 | 0.00 | 0.00 | 0.00 | 0.00 | 0.00 |
| 32 | PubExpHealth | 0.00 | 0.00 | 0.00 | 0.00 | 0.00 | 0.00 | 0.00 |
| 33 | PublicExp Administration | 0.00 | 0.00 | 0.00 | 0.00 | 0.00 | 0.00 | 0.00 |
| 34 | PublicExpPub Order & Defense | 0.00 | 0.00 | 0.00 | 0.00 | 0.00 | 0.00 | 0.00 |
| 35 | PunExp Other Public Admin | 0.00 | 0.00 | 0.00 | 0.00 | 0.00 | 0.00 | 0.00 |
| 36 | PubExp Household Tran | 0.00 | 0.00 | 0.00 | 0.00 | 0.00 | 0.00 | 0.00 |
| 37 | PubInvAgric & Rural Development | 0.00 | 0.00 | 0.00 | 0.00 | 0.00 | 0.00 | 0.00 |
| 38 | PubInvIndustry | 0.00 | 0.00 | 0.00 | 0.00 | 0.00 | 0.00 | 0.00 |
| 39 | PubInvTrade | 0.00 | 0.00 | 0.00 | 0.00 | 0.00 | 0.00 | 0.00 |
| 40 | PubInv Transportation & Communication | 0.00 | 0.00 | 0.00 | 0.00 | 0.00 | 0.00 | 0.00 |
| 41 | PubInvEduc&health | 0.00 | 0.00 | 0.00 | 0.00 | 0.00 | 0.00 | 0.00 |
| 42 | PubInv Administration | 0.00 | 0.00 | 0.00 | 0.00 | 0.00 | 0.00 | 0.00 |
| 43 | PubInvOthers | 0.00 | 0.00 | 0.00 | 0.00 | 0.00 | 0.00 | 0.00 |
| 44 | Commodities Taxes (Domestic) | 0.00 | 2,307.05 | 285.40 | 106.95 | 27.21 | 2,421.49 | 1,844.62 |
| 45 | Commodities Taxes (Imports) | 0.00 | 441.72 | 54.64 | 20.48 | 5.21 | 463.63 | 353.18 |
| 46 | Public Current | 0.00 | 1,273.11 | 737.33 | 430.05 | 79.35 | 1,402.44 | 1,595.92 |
| 47 | Public Capital | 0.00 | 0.00 | 0.00 | 0.00 | 0.00 | 0.00 | 0.00 |
| 48 | Private Capital | 0.00 | 3,031.02 | 2,962.95 | 436.81 | 6.76 | 5,567.89 | 13,134.64 |
| 49 | Changes in inventories | 0.00 | 0.00 | 0.00 | 0.00 | 0.00 | 0.00 | 0.00 |
| 50 | ROW current | 0.00 | 6,700.82 | 828.94 | 310.63 | 79.02 | 7,033.20 | 5,357.70 |
| 51 | ROW capital | 0.00 | 0.00 | 0.00 | 0.00 | 0.00 | 0.00 | 0.00 |
| | TOTAL | 345,270.11 | 35,848.78 | 7,540.68 | 2,808.47 | 414.62 | 52,140.76 | 64,608.16 |

Continue

| | | 26 | 27 | 28 | 29 | 30 | 31 | 32 |
|---|---|---|---|---|---|---|---|---|
| | | Urban Indian expenditure | Urban Others expenditure | Non-citizen expenditure | Companies expenditure | PubExp Agriculture | PubExp Education | PubExp Health |
| 1 | Agriculture and livestock | 299.43 | 3.46 | 473.41 | 0.00 | 1,323.00 | 0.00 | 0.00 |
| 2 | Forestry and logging products | 0.00 | 0.00 | 0.00 | 0.00 | 0.00 | 0.00 | 0.00 |
| 3 | Fish etc. | 26.28 | 0.21 | 1,358.10 | 0.00 | 0.00 | 0.00 | 0.00 |
| 4 | Mining & quarrying | 0.00 | 0.00 | 0.00 | 0.00 | 0.00 | 0.00 | 0.00 |
| 5 | Manufacturing | 3,304.72 | 142.84 | 2,219.82 | 0.00 | 0.00 | 0.00 | 0.00 |
| 6 | Electricity, gas & water | 489.74 | 52.10 | 27.99 | 0.00 | 0.00 | 0.00 | 0.00 |
| 7 | Building and construcions | 17.95 | 2.72 | 19.19 | 0.00 | 0.00 | 0.00 | 0.00 |
| 8 | Wholesale and retail trade | 132.95 | 7.73 | 87.78 | 0.00 | 0.00 | 0.00 | 0.00 |
| 9 | Hotel& restaurant | 1,058.77 | 127.10 | 361.86 | 0.00 | 0.00 | 0.00 | 0.00 |
| 10 | Transport and communication | 1,394.68 | 101.81 | 253.15 | 0.00 | 0.00 | 0.00 | 0.00 |
| 11 | Financial, insurance and real estate | 1,764.98 | 101.70 | 432.39 | 0.00 | 0.00 | 0.00 | 0.00 |
| 12 | Business services | 83.91 | 4.42 | 26.88 | 0.00 | 0.00 | 0.00 | 0.00 |
| 13 | Education | 102.11 | 9.77 | 44.96 | 0.00 | 0.00 | 11,335.99 | 0.00 |
| 14 | Health | 324.52 | 11.21 | 9.05 | 0.00 | 0.00 | 0.00 | 4,198.41 |
| 15 | Other private services | 349.76 | 54.51 | 127.28 | 0.00 | 0.00 | 0.00 | 0.00 |
| 16 | General administration | 0.00 | 0.00 | 0.00 | 0.00 | 0.00 | 0.00 | 0.00 |
| 17 | Public order and defence | 0.00 | 0.00 | 0.00 | 0.00 | 0.00 | 0.00 | 0.00 |
| 18 | Other Public Administration | 0.00 | 0.00 | 0.00 | 0.00 | 0.00 | 0.00 | 0.00 |
| 19 | Factor of production | 0.00 | 0.00 | 0.00 | 0.00 | 0.00 | 0.00 | 0.00 |
| 20 | Rural Malay | 0.00 | 0.00 | 0.00 | 3,638.28 | 0.00 | 0.00 | 0.00 |
| 21 | Rural Chinese | 0.00 | 0.00 | 0.00 | 450.08 | 0.00 | 0.00 | 0.00 |
| 22 | Rural Indian | 214.94 | 0.00 | 0.00 | 168.66 | 0.00 | 0.00 | 0.00 |
| 23 | Rural Others | 0.00 | 30.02 | 0.00 | 42.90 | 0.00 | 0.00 | 0.00 |
| 24 | Urban Malay | 0.00 | 0.00 | 0.00 | 5,878.66 | 0.00 | 0.00 | 0.00 |
| 25 | Urban Chinese | 0.00 | 0.00 | 0.00 | 7,159.63 | 0.00 | 0.00 | 0.00 |
| 26 | Urban Indian | 430.46 | 0.00 | 0.00 | 872.17 | 0.00 | 0.00 | 0.00 |
| 27 | Urban others | 0.00 | 172.49 | 0.00 | 109.52 | 0.00 | 0.00 | 0.00 |
| 28 | Non-citizen | 0.00 | 0.00 | 81.46 | 4,414.10 | 0.00 | 0.00 | 0.00 |
| 29 | Companies | 0.00 | 0.00 | 0.00 | 0.00 | 0.00 | 0.00 | 0.00 |
| 30 | PubExpAgriculture | 0.00 | 0.00 | 0.00 | 0.00 | 0.00 | 0.00 | 0.00 |
| 31 | PubExpEducation | 0.00 | 0.00 | 0.00 | 0.00 | 0.00 | 0.00 | 0.00 |
| 32 | PubExpHealth | 0.00 | 0.00 | 0.00 | 0.00 | 0.00 | 0.00 | 0.00 |
| 33 | PublicExp Administration | 0.00 | 0.00 | 0.00 | 0.00 | 0.00 | 0.00 | 0.00 |
| 34 | PublicExpPub Order & Defense | 0.00 | 0.00 | 0.00 | 0.00 | 0.00 | 0.00 | 0.00 |
| 35 | PunExpOther Public Admin | 0.00 | 0.00 | 0.00 | 0.00 | 0.00 | 0.00 | 0.00 |
| 36 | PubExp Household Tran | 0.00 | 0.00 | 0.00 | 0.00 | 0.00 | 0.00 | 0.00 |
| 37 | PubInvAgric & RuralDevelopment | 0.00 | 0.00 | 0.00 | 0.00 | 0.00 | 0.00 | 0.00 |
| 38 | PubInvIndustry | 0.00 | 0.00 | 0.00 | 0.00 | 0.00 | 0.00 | 0.00 |
| 39 | PubInvTrade | 0.00 | 0.00 | 0.00 | 0.00 | 0.00 | 0.00 | 0.00 |
| 40 | PubInvTransportation& Communication | 0.00 | 0.00 | 0.00 | 0.00 | 0.00 | 0.00 | 0.00 |
| 41 | PubInvEduc&health | 0.00 | 0.00 | 0.00 | 0.00 | 0.00 | 0.00 | 0.00 |
| 42 | PubInvAdministration | 0.00 | 0.00 | 0.00 | 0.00 | 0.00 | 0.00 | 0.00 |
| 43 | PubInvOthers | 0.00 | 0.00 | 0.00 | 0.00 | 0.00 | 0.00 | 0.00 |
| 44 | Commodities Taxes (Domestic) | 470.11 | 41.80 | 227.52 | 134.00 | 0.00 | 0.00 | 0.00 |
| 45 | Commodities Taxes (Imports) | 90.01 | 8.00 | 43.56 | 166.30 | 0.00 | 0.00 | 0.00 |
| 46 | Public Current | 616.83 | 134.53 | 745.43 | 27,262.66 | 0.00 | 0.00 | 0.00 |
| 47 | Public Capital | 0.00 | 0.00 | 0.00 | 0.00 | 0.00 | 0.00 | 0.00 |
| 48 | Private Capital | 767.36 | 10.64 | 3,012.46 | 72,593.51 | 0.00 | 0.00 | 0.00 |
| 49 | Changes in inventories | 0.00 | 0.00 | 0.00 | 0.00 | 0.00 | 0.00 | 0.00 |
| 50 | ROW current | 1,365.44 | 121.41 | 660.83 | 46,494.00 | 0.00 | 0.00 | 0.00 |
| 51 | ROW capital | 0.00 | 0.00 | 0.00 | 32,257.00 | 0.00 | 0.00 | 0.00 |
| | TOTAL | 13,304.95 | 1,138.47 | 10,213.14 | 201,641.47 | 1,323.00 | 11,335.99 | 4,198.41 |

Continue

| | | 33 | 34 | 35 | 36 | 37 | 38 | 39 |
|---|---|---|---|---|---|---|---|---|
| | | PubExp GenAdmin | PublicExp Order & Defense | PubExp Others | PublicExp Transfer | PubInv Agri & RurDev | PubInv industry | PubInv Trade |
| 1 | Agriculture and livestock | 0.00 | 0.00 | 0.00 | 0.00 | 328.32 | 0.00 | 0.00 |
| 2 | Forestry and logging products | 0.00 | 0.00 | 0.00 | 0.00 | 13.68 | 0.00 | 0.00 |
| 3 | Fish etc. | 0.00 | 0.00 | 0.00 | 0.00 | 34.77 | 0.00 | 0.00 |
| 4 | Mining & quarrying | 0.00 | 0.00 | 0.00 | 0.00 | 0.00 | 0.00 | 0.00 |
| 5 | Manufacturing | 0.00 | 0.00 | 0.00 | 0.00 | 0.00 | 1,679.79 | 0.00 |
| 6 | Electricity, gas & water | 0.00 | 0.00 | 0.00 | 0.00 | 0.00 | 0.00 | 0.00 |
| 7 | Building and construcions | 0.00 | 0.00 | 0.00 | 0.00 | 0.00 | 0.00 | 0.00 |
| 8 | Wholesale and retail trade | 0.00 | 0.00 | 0.00 | 0.00 | 0.00 | 0.00 | 626.43 |
| 9 | Hotel& restaurant | 0.00 | 0.00 | 0.00 | 0.00 | 0.00 | 0.00 | 0.00 |
| 10 | Transport and communication | 0.00 | 0.00 | 0.00 | 0.00 | 0.00 | 0.00 | 0.00 |
| 11 | Financial, insurance and real estate | 0.00 | 0.00 | 0.00 | 0.00 | 0.00 | 0.00 | 0.00 |
| 12 | Business services | 0.00 | 0.00 | 0.00 | 0.00 | 0.00 | 0.00 | 0.00 |
| 13 | Education | 0.00 | 0.00 | 0.00 | 0.00 | 0.00 | 0.00 | 0.00 |
| 14 | Health | 0.00 | 0.00 | 0.00 | 0.00 | 0.00 | 0.00 | 0.00 |
| 15 | Other private services | 0.00 | 0.00 | 0.00 | 0.00 | 0.00 | 0.00 | 0.00 |
| 16 | General administration | 5,055.72 | 0.00 | 0.00 | 0.00 | 0.00 | 0.00 | 0.00 |
| 17 | Public order and defence | 0.00 | 8,504.26 | 0.00 | 0.00 | 0.00 | 0.00 | 0.00 |
| 18 | Other Public Administration | 0.00 | | 4,444.50 | 0.00 | 0.00 | 0.00 | 0.00 |
| 19 | Factor of production | 0.00 | 0.00 | 0.00 | 0.00 | 0.00 | 0.00 | 0.00 |
| 20 | Rural Malay | 0.00 | 0.00 | 0.00 | 2,416.61 | 0.00 | 0.00 | 0.00 |
| 21 | Rural Chinese | 0.00 | 0.00 | 0.00 | 298.95 | 0.00 | 0.00 | 0.00 |
| 22 | Rural Indian | 0.00 | 0.00 | 0.00 | 112.03 | 0.00 | 0.00 | 0.00 |
| 23 | Rural Others | 0.00 | 0.00 | 0.00 | 28.50 | 0.00 | 0.00 | 0.00 |
| 24 | Urban Malay | 0.00 | 0.00 | 0.00 | 2,536.48 | 0.00 | 0.00 | 0.00 |
| 25 | Urban Chinese | 0.00 | 0.00 | 0.00 | 1,932.22 | 0.00 | 0.00 | 0.00 |
| 26 | Urban Indian | 0.00 | 0.00 | 0.00 | 492.44 | 0.00 | 0.00 | 0.00 |
| 27 | Urban others | 0.00 | 0.00 | 0.00 | 43.79 | 0.00 | 0.00 | 0.00 |
| 28 | Non-citizen | 0.00 | 0.00 | 0.00 | 0.00 | 0.00 | 0.00 | 0.00 |
| 29 | Companies | 0.00 | 0.00 | 0.00 | 0.00 | 0.00 | 0.00 | 0.00 |
| 30 | PubExpAgriculture | 0.00 | 0.00 | 0.00 | 0.00 | 0.00 | 0.00 | 0.00 |
| 31 | PubExpEducation | 0.00 | 0.00 | 0.00 | 0.00 | 0.00 | 0.00 | 0.00 |
| 32 | PubExpHealth | 0.00 | 0.00 | 0.00 | 0.00 | 0.00 | 0.00 | 0.00 |
| 33 | PublicExp Administration | 0.00 | 0.00 | 0.00 | 0.00 | 0.00 | 0.00 | 0.00 |
| 34 | PublicExpPub Order & Defense | 0.00 | 0.00 | 0.00 | 0.00 | 0.00 | 0.00 | 0.00 |
| 35 | PunExpOther Public Admin | 0.00 | 0.00 | 0.00 | 0.00 | 0.00 | 0.00 | 0.00 |
| 36 | PubExp Household Tran | 0.00 | 0.00 | 0.00 | 0.00 | 0.00 | 0.00 | 0.00 |
| 37 | PubInvAgric & Rural Development | 0.00 | 0.00 | 0.00 | 0.00 | 0.00 | 0.00 | 0.00 |
| 38 | PubInvIndustry | 0.00 | 0.00 | 0.00 | 0.00 | 0.00 | 0.00 | 0.00 |
| 39 | PubInvTrade | 0.00 | 0.00 | 0.00 | 0.00 | 0.00 | 0.00 | 0.00 |
| 40 | PubInvTransportation & Communication | 0.00 | 0.00 | 0.00 | 0.00 | 0.00 | 0.00 | 0.00 |
| 41 | PubInvEduc&health | 0.00 | 0.00 | 0.00 | 0.00 | 0.00 | 0.00 | 0.00 |
| 42 | PubInvAdministration | 0.00 | 0.00 | 0.00 | 0.00 | 0.00 | 0.00 | 0.00 |
| 43 | PubInvOthers | 0.00 | 0.00 | 0.00 | 0.00 | 0.00 | 0.00 | 0.00 |
| 44 | Commodities Taxes (Domestic) | 0.00 | 0.00 | 0.00 | 0.00 | 0.00 | 0.00 | 0.00 |
| 45 | Commodities Taxes (Imports) | 0.00 | 0.00 | 0.00 | 0.00 | 0.00 | 0.00 | 0.00 |
| 46 | Public Current | 0.00 | 0.00 | 0.00 | 0.00 | 0.00 | 0.00 | 0.00 |
| 47 | Public Capital | 0.00 | 0.00 | 0.00 | 0.00 | 0.00 | 0.00 | 0.00 |
| 48 | Private Capital | 0.00 | 0.00 | 0.00 | 0.00 | 0.00 | 0.00 | 0.00 |
| 49 | Changes in inventories | 0.00 | 0.00 | 0.00 | 0.00 | 0.00 | 0.00 | 0.00 |
| 50 | ROW current | 0.00 | 0.00 | 0.00 | 0.00 | 0.00 | 0.00 | 0.00 |
| 51 | ROW capital | 0.00 | 0.00 | 0.00 | 0.00 | 0.00 | 0.00 | 0.00 |
| | TOTAL | 5,055.72 | 8,504.26 | 4,444.50 | 7,861.00 | 376.77 | 1,679.79 | 626.43 |

Continue

| | | 40 | 41 | 42 | 43 | 44 | 45 | 46 |
|---|---|---|---|---|---|---|---|---|
| | | PubInv Transp & com | PubInv Educ & health | PubInv Admin | Pub.Inv Other | ComTaxes (domestic) | Commdtaxes (import) | Pub Current |
| 1 | Agriculture and livestock | 0.00 | 0.00 | 0.00 | 0.00 | 0.00 | 0.00 | 0.00 |
| 2 | Forestry and logging products | 0.00 | 0.00 | 0.00 | 0.00 | 0.00 | 0.00 | 0.00 |
| 3 | Fish etc. | 0.00 | 0.00 | 0.00 | 0.00 | 0.00 | 0.00 | 0.00 |
| 4 | Mining & quarrying | 0.00 | 0.00 | 0.00 | 0.00 | 0.00 | 0.00 | 0.00 |
| 5 | Manufacturing | 0.00 | 0.00 | 0.00 | 0.00 | 0.00 | 0.00 | 0.00 |
| 6 | Electricity, gas & water | 0.00 | 0.00 | 0.00 | 0.00 | 0.00 | 0.00 | 0.00 |
| 7 | Building and construcions | 0.00 | 0.00 | 0.00 | 0.00 | 0.00 | 0.00 | 0.00 |
| 8 | Wholesale and retail trade | 0.00 | 0.00 | 0.00 | 0.00 | 0.00 | 0.00 | 0.00 |
| 9 | Hotel& restaurant | 0.00 | 0.00 | 0.00 | 0.00 | 0.00 | 0.00 | 0.00 |
| 10 | Transport and communication | 847.59 | 0.00 | 0.00 | 0.00 | 0.00 | 0.00 | 0.00 |
| 11 | Financial, insurance and real estate | 0.00 | 0.00 | 0.00 | 0.00 | 0.00 | 0.00 | 0.00 |
| 12 | Business services | 0.00 | 0.00 | 0.00 | 0.00 | 0.00 | 0.00 | 0.00 |
| 13 | Education | 0.00 | 2,432.19 | 0.00 | 0.00 | 0.00 | 0.00 | 0.00 |
| 14 | Health | 0.00 | 851.01 | 0.00 | 0.00 | 0.00 | 0.00 | 0.00 |
| 15 | Other private services | 0.00 | 0.00 | 0.00 | 235.98 | 0.00 | 0.00 | 0.00 |
| 16 | General administration | 0.00 | 0.00 | 2,215.02 | 0.00 | 0.00 | 0.00 | 0.00 |
| 17 | Public order and defence | 0.00 | 0.00 | 795.15 | 0.00 | 0.00 | 0.00 | 0.00 |
| 18 | Other Public Administration | 0.00 | 0.00 | 0.00 | 536.94 | 0.00 | 0.00 | 0.00 |
| 19 | Factor of production | 0.00 | 0.00 | 0.00 | 0.00 | 0.00 | 0.00 | 0.00 |
| 20 | Rural Malay | 0.00 | 0.00 | 0.00 | 0.00 | 0.00 | 0.00 | 0.00 |
| 21 | Rural Chinese | 0.00 | 0.00 | 0.00 | 0.00 | 0.00 | 0.00 | 0.00 |
| 22 | Rural Indian | 0.00 | 0.00 | 0.00 | 0.00 | 0.00 | 0.00 | 0.00 |
| 23 | Rural Others | 0.00 | 0.00 | 0.00 | 0.00 | 0.00 | 0.00 | 0.00 |
| 24 | Urban Malay | 0.00 | 0.00 | 0.00 | 0.00 | 0.00 | 0.00 | 0.00 |
| 25 | Urban Chinese | 0.00 | 0.00 | 0.00 | 0.00 | 0.00 | 0.00 | 0.00 |
| 26 | Urban Indian | 0.00 | 0.00 | 0.00 | 0.00 | 0.00 | 0.00 | 0.00 |
| 27 | Urban others | 0.00 | 0.00 | 0.00 | 0.00 | 0.00 | 0.00 | 0.00 |
| 28 | Non-citizen | 0.00 | 0.00 | 0.00 | 0.00 | 0.00 | 0.00 | 0.00 |
| 29 | Companies | 0.00 | 0.00 | 0.00 | 0.00 | 0.00 | 0.00 | 0.00 |
| 30 | PubExpAgriculture | 0.00 | 0.00 | 0.00 | 0.00 | 0.00 | 0.00 | 1,323.00 |
| 31 | PubExpEducation | 0.00 | 0.00 | 0.00 | 0.00 | 0.00 | 0.00 | 11,335.99 |
| 32 | PubExpHealth | 0.00 | 0.00 | 0.00 | 0.00 | 0.00 | 0.00 | 4,198.41 |
| 33 | PublicExp Administration | 0.00 | 0.00 | 0.00 | 0.00 | 0.00 | 0.00 | 5,055.72 |
| 34 | PublicExpPub Order & Defense | 0.00 | 0.00 | 0.00 | 0.00 | 0.00 | 0.00 | 8,504.26 |
| 35 | PunExpOther Public Admin | 0.00 | 0.00 | 0.00 | 0.00 | 0.00 | 0.00 | 4,444.50 |
| 36 | PubExp Household Tran | 0.00 | 0.00 | 0.00 | 0.00 | 0.00 | 0.00 | 7,861.00 |
| 37 | PubInvAgric & Rural Development | 0.00 | 0.00 | 0.00 | 0.00 | 0.00 | 0.00 | 0.00 |
| 38 | PubInvIndustry | 0.00 | 0.00 | 0.00 | 0.00 | 0.00 | 0.00 | 0.00 |
| 39 | PubInvTrade | 0.00 | 0.00 | 0.00 | 0.00 | 0.00 | 0.00 | 0.00 |
| 40 | PubInvTransportation & Communication | 0.00 | 0.00 | 0.00 | 0.00 | 0.00 | 0.00 | 0.00 |
| 41 | PubInvEduc&health | 0.00 | 0.00 | 0.00 | 0.00 | 0.00 | 0.00 | 0.00 |
| 42 | PubInvAdministration | 0.00 | 0.00 | 0.00 | 0.00 | 0.00 | 0.00 | 0.00 |
| 43 | PubInvOthers | 0.00 | 0.00 | 0.00 | 0.00 | 0.00 | 0.00 | 0.00 |
| 44 | Commodities Taxes (Domestic) | 0.00 | 0.00 | 0.00 | 0.00 | 0.00 | 0.00 | 0.00 |
| 45 | Commodities Taxes (Imports) | 0.00 | 0.00 | 0.00 | 0.00 | 0.00 | 0.00 | 0.00 |
| 46 | Public Current | 0.00 | 0.00 | 0.00 | 0.00 | 14,650.40 | 5,826.82 | 0.00 |
| 47 | Public Capital | 0.00 | 0.00 | 0.00 | 0.00 | 0.00 | 0.00 | 11,557.00 |
| 48 | Private Capital | 0.00 | 0.00 | 0.00 | 0.00 | 0.00 | 0.00 | 0.00 |
| 49 | Changes in inventories | 0.00 | 0.00 | 0.00 | 0.00 | 0.00 | 0.00 | 0.00 |
| 50 | ROW current | 0.00 | 0.00 | 0.00 | 0.00 | 0.00 | 0.00 | 919.00 |
| 51 | ROW capital | 0.00 | 0.00 | 0.00 | 0.00 | 0.00 | 0.00 | 0.00 |
| | TOTAL | 847.59 | 3,283.20 | 3,010.17 | 772.92 | 14,650.40 | 5,826.82 | 55,198.88 |

Continue

| | | 47 | 48 | 49 | 50 | 51 | |
|---|---|---|---|---|---|---|---|
| | | Pub Capital | Private Capital | Changes in inventories | ROW current | ROW capital | TOTAL |
| 1 | Agriculture and livestock | 0.00 | 1,337.79 | -1,738.21 | 2,451.93 | 0.00 | 20,908.05 |
| 2 | Forestry and logging products | 0.00 | 231.42 | 924.71 | 2,567.86 | 0.00 | 11,190.68 |
| 3 | Fish etc. | 0.00 | 301.53 | -332.19 | 132.88 | 0.00 | 5,452.83 |
| 4 | Mining & quarrying | 0.00 | 2,896.17 | 498.40 | 25,416.86 | 0.00 | 43,857.89 |
| 5 | Manufacturing | 0.00 | 19,197.00 | -16,560.60 | 322,318.51 | 0.00 | 495,306.31 |
| 6 | Electricity, gas & water | 0.00 | 5,135.70 | -5,135.68 | 5.31 | 0.00 | 17,380.11 |
| 7 | Building and construcions | 0.00 | 592.23 | 38,750.22 | 1,950.05 | 0.00 | 45,091.03 |
| 8 | Wholesale and retail trade | 0.00 | 315.78 | 1,691.61 | 9,451.17 | 0.00 | 52,347.60 |
| 9 | Hotel& restaurant | 0.00 | 311.22 | -311.22 | 0.00 | 0.00 | 20,904.69 |
| 10 | Transport and communication | 0.00 | 7,435.65 | -8,100.34 | 20,211.77 | 0.00 | 52,043.73 |
| 11 | Financial, insurance and real estate | 0.00 | 1,422.22 | 20,219.24 | 2,920.36 | 0.00 | 60,890.76 |
| 12 | Business services | 0.00 | 885.78 | -581.77 | 10,639.05 | 0.00 | 21,701.45 |
| 13 | Education | 0.00 | 0.00 | -2,432.19 | 150.06 | 0.00 | 13,468.99 |
| 14 | Health | 0.00 | 0.00 | -851.01 | 209.61 | 0.00 | 7,316.64 |
| 15 | Other private services | 0.00 | 721.62 | 373.52 | 952.86 | 0.00 | 10,456.95 |
| 16 | General administration | 0.00 | 0.00 | -2,215.02 | 0.00 | 0.00 | 5,456.14 |
| 17 | Public order and defence | 0.00 | 0.00 | -794.89 | 0.00 | 0.00 | 8,540.98 |
| 18 | Other Public Administration | 0.00 | 0.00 | -481.66 | 1.15 | 0.00 | 4,513.13 |
| 19 | Factor of production | 0.00 | 0.00 | 0.00 | 0.00 | 0.00 | 345,270.11 |
| 20 | Rural Malay | 0.00 | 0.00 | 0.00 | 0.00 | 0.00 | 35,848.78 |
| 21 | Rural Chinese | 0.00 | 0.00 | 0.00 | 0.00 | 0.00 | 7,540.68 |
| 22 | Rural Indian | 0.00 | 0.00 | 0.00 | 0.00 | 0.00 | 2,808.48 |
| 23 | Rural Others | 0.00 | 0.00 | 0.00 | 0.00 | 0.00 | 414.63 |
| 24 | Urban Malay | 0.00 | 0.00 | 0.00 | 325.30 | 0.00 | 52,140.76 |
| 25 | Urban Chinese | 0.00 | 0.00 | 0.00 | 325.30 | 0.00 | 64,608.17 |
| 26 | Urban Indian | 0.00 | 0.00 | 0.00 | 325.30 | 0.00 | 13,304.95 |
| 27 | Urban others | 0.00 | 0.00 | 0.00 | 325.30 | 0.00 | 1,138.47 |
| 28 | Non-citizen | 0.00 | 0.00 | 0.00 | 0.00 | 0.00 | 10,213.13 |
| 29 | Companies | 0.00 | 0.00 | 0.00 | 8,674.01 | 0.00 | 201,641.48 |
| 30 | PubExpAgriculture | 0.00 | 0.00 | 0.00 | 0.00 | 0.00 | 1,323.00 |
| 31 | PubExpEducation | 0.00 | 0.00 | 0.00 | 0.00 | 0.00 | 11,335.99 |
| 32 | PubExpHealth | 0.00 | 0.00 | 0.00 | 0.00 | 0.00 | 4,198.41 |
| 33 | PublicExp Administration | 0.00 | 0.00 | 0.00 | 0.00 | 0.00 | 5,055.72 |
| 34 | PublicExpPub Order & Defense | 0.00 | 0.00 | 0.00 | 0.00 | 0.00 | 8,504.26 |
| 35 | PunExpOther Public Admin | 0.00 | 0.00 | 0.00 | 0.00 | 0.00 | 4,444.50 |
| 36 | PubExp Household Tran | 0.00 | 0.00 | 0.00 | 0.00 | 0.00 | 7,861.00 |
| 37 | PubInvAgric & Rural Development | 376.77 | 0.00 | 0.00 | 0.00 | 0.00 | 376.77 |
| 38 | PubInvIndustry | 1,679.79 | 0.00 | 0.00 | 0.00 | 0.00 | 1,679.79 |
| 39 | PubInvTrade | 626.43 | 0.00 | 0.00 | 0.00 | 0.00 | 626.43 |
| 40 | PubInvTransportation & Communication | 847.59 | 0.00 | 0.00 | 0.00 | 0.00 | 847.59 |
| 41 | PubInvEduc&health | 3,283.20 | 0.00 | 0.00 | 0.00 | 0.00 | 3,283.20 |
| 42 | PubInvAdministration | 3,010.17 | 0.00 | 0.00 | 0.00 | 0.00 | 3,010.17 |
| 43 | PubInvOthers | 772.92 | 0.00 | 0.00 | 0.00 | 0.00 | 772.92 |
| 44 | Commodities Taxes (Domestic) | 96.87 | 372.83 | 0.00 | 1,082.05 | 0.00 | 14,650.40 |
| 45 | Commodities Taxes (Imports) | 206.59 | 795.18 | 0.00 | 4.20 | 0.00 | 5,826.82 |
| 46 | Public Current | 0.00 | 0.00 | 0.00 | 444.00 | 0.00 | 55,198.88 |
| 47 | Public Capital | 0.00 | 10,897.14 | 0.00 | 0.00 | 864.00 | 23,318.14 |
| 48 | Private Capital | 0.00 | 0.00 | 0.00 | 0.00 | -881.00 | 100,643.04 |
| 49 | Changes in inventories | 4,727.58 | 18,195.33 | 0.00 | 0.00 | 0.00 | 22,922.91 |
| 50 | ROW current | 7,690.23 | 29,598.46 | 0.00 | 26,559.00 | 0.00 | 405,169.63 |
| 51 | ROW capital | 0.00 | 0.00 | 0.00 | -32,274.00 | 0.00 | -17.00 |
| | TOTAL | 23,318.14 | 100,643.05 | 22,922.91 | 405,169.87 | -17.00 | 2,312,790.15 |